\documentclass[11pt,prd,letterpaper,tightenlines,superscriptaddress,showpacs,psfig]{revtex4}
\usepackage{amsmath,amssymb,amsfonts,graphicx,dcolumn}
\usepackage{epstopdf}
\usepackage{color}
\usepackage[utf8]{inputenc}
\usepackage{amsthm}
\usepackage{fontenc}
\usepackage{slashed}
\usepackage{hyperref}
\usepackage{bm}
\usepackage{graphics}
\RequirePackage{slashed}
\usepackage{cancel}
\usepackage[normalem]{ulem}

\usepackage{array,booktabs}
\newcolumntype{C}{>{$\displaystyle}c<{$}}

\begin{document}


\title{Four-jet production via double parton scattering in $pA$ collisions at the LHC}

\author{Boris Blok}
\email{blok@physics.technion.ac.il}
\affiliation{ Department of Physics, Technion, Israel Institute of Technology, Haifa, 32000 Israel}
\author{Federico Alberto Ceccopieri}
\email{federico.ceccopieri@hotmail.it}	
\affiliation{ Department of Physics, Technion, Israel Institute of Technology, Haifa, 32000 Israel}
\affiliation{IFPA, Université de Li\`ege, B4000, Li\`ege, Belgium}

\begin{abstract}
\vspace{0.5cm}
\indent We present predictions for the double parton scattering (DPS)
four-jet production cross sections in $pA$ collisions at the LHC.
Relying on the experimental capabilities to 
correlate centrality with impact parameter $B$ of 
the proton-nucleus collision, 
we discuss a strategy to extract the double  parton scattering contributions in $pA$ collisions,
which gives direct access to double parton distribution in the nucleon.
We show that the production cross sections via DPS of 
four jets, out of which two may be light- or heavy-quark jets,
are large enough to allow the method to be used already with data accumulated in 2016 $pA$ run.
 \end{abstract}

\maketitle

\section{Introduction}
\label{Sec:Intro}
\par The flux of incoming partons in hadron-induced reactions increases with the collision energy so that multiple parton interactions (MPI) take place,
both in $pp$ and $pA$ collisions. The study of MPIs started in eighties in Tevatron era~\cite{TreleaniPaver82,Paver:1983hi,mufti}, both experimentally  and theoretically.
Recently a significant progress was achieved in the study of MPI, in particular of double parton scattering (DPS). From the theoretical point  of view a new self consistent 
pQCD based formalism was developed both for 
$pp$~\cite{stirling,BDFS1,Diehl,stirling1,BDFS2,Diehl2,BDFS3,BDFS4,Diehl:2017kgu,Manohar:2012jr} and $pA$ DPS collisions~\cite{BSW} (see~\cite{book} for recent reviews). Recent observations of double open charm ~\cite{Belyaev,LHCb,LHCb1,LHCb2} and 
same sign $WW$ ($ssWW$) production~\cite{Sirunyan:2019zox} clearly show the   existence of DPS interactions in $pp$ collisions. 
\par  The MPI interactions play a major role in the Underlying Event (UE) and thus are taken into account in all MC generators developed for the LHC~\cite{pythia,herwig}.
On the other hand the study of DPS will lead to understanding of two parton correlations  in the nucleon. In particular 
the DPS cross sections involve new non-perturbative two-body  quantities, the so-called two particle Generalised  Parton Distribution Functions ($_2$GPDs), which encode novel  features of the non-perturbative nucleon structure.
Such distributions have the potential to unveil two-parton correlations in the nucleon structure~\cite{calucci,Rinaldi:2018slz} and to give access to information complementary to the one obtained from nucleon one-body distributions. 

\par The study of MPI and in particular of the DPS reactions in  $pA$ collisions is important for our understanding of  MPI in $pp$ collisions and it constitutes a  benchmark of the theoretical formalism available for these processes. On the other hand the MPI in $pA$ collisions may play an important role in underlying event (UE) and high multiplicity events in $pA$ collisions. Moreover it was argued in Ref.~\cite{BSW} that they are directly related to longitudinal parton correlations in the nucleon.
\par The theory of MPI and in particular DPS in $pA$ collisions was first developed in~\cite{Strikman:2001gz}, where it was shown that there are two DPS contributions at work in such a case. 
\begin{figure}[t]~
\includegraphics[scale=0.8]{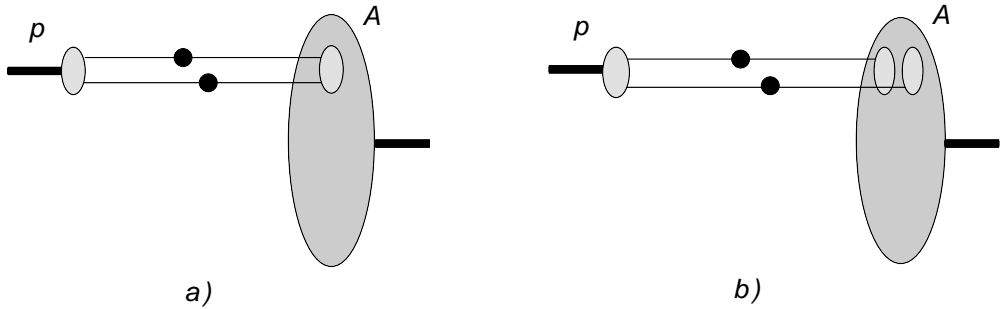}
\caption{\textsl{Pictorial representation of DPS process in $pA$ collisions via a) DPS1 and b) DPS2 mechanisms. The light grey blobs indicate nucleons, darker grey ones the nucleus and black ones the hard interactions.}}
\label{plot:DPS_pictorial}
\end{figure}
First, there is the so-called DPS1 contribution, depicted in the left panel of Fig.~(\ref{plot:DPS_pictorial}), in which two partons from the incoming nucleon interact with two partons in the target nucleon in the nucleus, making 
such a process formally identical to DPS in the $pp$ collisions. Next there is a new type of contribution, 
depicted in the right panel of Fig.~(\ref{plot:DPS_pictorial})
and often called DPS2, in which 
two partons from the incoming nucleon interact with two partons  each of them belonging to the distinct nucleons in the target nucleus located at the same impact parameter. Such a contribution is parametrically enhanced by a factor $A^{1/3}$ over the DPS1 contribution, $A$ being the atomic number of the nucleus.
\par In the recent past a number of theoretical studies  have appeared which 
focus on the study of DPS contributions in $pA$ collisions~\cite{Helenius:2019uge,sde,Cattaruzza:2004qb,BSW,Calucci:2013pza,Fedkevych:2019ofc}.
However, although many interesting theoretical studies of  DPS2 were performed recently, the problem remains is how to observe DPS2 experimentally. The main issue 
is obviously the large SPS (leading twist) background in such processes, that makes the observation of the DPS contributions, which are next to leading twist phenomena, a rather complicated task. 
\par  Recently however a new method was suggested  in Ref.~\cite{Alvioli:2019kcy}, which allows to separate DPS2 from the leading twist (and DPS1) contributions.  The method  exploits the  
different dependence on the impact parameter $B$
on the various contributions to pA cross section for a given final state:
while the SPS and DPS1 contributions are  proportional to  the nuclear thickness function $T(B)$, the  DPS2 one is proportional to the square of $T(B)$. Therefore the cross section producing a given final state can be schematically written as
\cite{Alvioli:2019kcy},:
\begin{equation}
\frac{d^2\sigma_{pA}}{d^2B}= 
\Big(\sigma^{LT}_{pA}+\sigma^{DPS1}_{pA}\Big) \frac{T(B)}{A} +\sigma_{pA}^{DPS2} \frac{T^2(B)}{\int d^2 B \, T^2(B)}\,,
\label{e1}
\end{equation}
where $T(B)$ is normalized to the atomic number $A$ of the nucleus. This approach was used in Ref.~\cite{Alvioli:2019kcy} to study two-dijets processes and, in Ref.~\cite{Blok:2019fgg}, to study processes involving the associated production of electroweak bosons and jets in $pA$ collisions. 
\par The latter strategy exploits the experimental capabilities to accurately relate centrality with the impact parameter $B$ of the $pA$ collisions.
The procedure for the determination of centrality in $pA$ collisions was  developed  \textsl{i.e.} by ATLAS~\cite{30}. It makes use of the measurement of the transverse energy $E_T$ 
deposited in the  pseudorapidity interval $-3.2\ge \eta  \ge   -4.9$ (i.e. along the nucleus direction) as a measure of centrality. It was shown in Ref.~\cite{35}     that        $E_T$ in this kinematics is not sensitive to production of hadrons at forward rapidities.  The $E_T$ distribution as a function of  the number of collisions $\nu$ (and thus on the impact parameter $B$) is presented in Refs.~\cite{25, 30, 35} (see also the related discussion in Ref.~\cite{Alvioli:2019kcy}).
\par The purpose of the present paper is to continue the research started in those works and pursue the emergence of DPS2 contribution in the four-jet final state. 
Indeed, the  observation of DPS in $pA$ collisions faces two main challenges: the first one, 
in common with DPS studies in $pp$ collisions, is tackling the large single parton scattering (SPS) background; the second one is given by the limited integrated luminosity accumulated in short $pA$ runs, which is  several orders of magnitude integrated lower than the one accumulated in $pp$ collisions.  Therefore the obviuous question is whether the number of observed DPS events is sufficient to overcome the systematic inaccuracy due to the  large SPS background.
Such question was studied for example in ~\cite{Blok:2019fgg} where we found that it is possible to separate SPS and DPS2 backgrounds for $Wjj$ final state.
\par The purpose of this paper is to 
investigate the possibility to isolate
the DPS2 contribution within multi-jet final state
and the necessary kinematic constraints.
We shall  calculate the cross sections as a function of impact parameter $B$ of the $pA$ collision,
for its various components  in both the four-jet ($4j$) and  two $b$-jet plus two light jets  ($2b2j$) final states and estimate the sensitivity to the DPS mechanisms for the considered final states. 
We shall see, that both these final states are the "golden plate" channel for the observation of the DPS2 mechanism.   
\par The paper is organised as follows. In Section~
\ref{Sec:theory} we review the theoretical formalism and the set up for our calculations. In Sec.~
\ref{4j} and ~\ref{2b2j} we analyze and discuss 
the results in the $4j$ and $2b2j$ final states, respectively. We summarize our results in conclusion.

\section{Theoretical Framework}
\label{Sec:theory}
\noindent
The cross section for the production of final states $C$ and $D$ 
in $pA$ collisions via double parton scattering can be written as the convolution of the double $_2$GPDs of the proton and the nucleus, $G_p$ and $G_A$, respectively~\cite{BSW,Strikman:2001gz}:
\begin{equation}
\frac{d \sigma^{CD}_{DPS} }{d\Omega_C d\Omega_D}=\int \frac{d^2\vec \Delta}{(2\pi)^2}\frac{d \hat \sigma_{ik}^C(x_1, x_3)}{d\Omega_C}\frac{d \hat \sigma_{jl}^{D}(x_2,x_4)}{d\Omega_D}G_p^{ij}(x_1,x_2, \vec \Delta )G_A^{kl}(x_3,x_4, -\vec \Delta)\,.
\label{Eq:DPS_momentum_space}
\end{equation}
Two parton GPDs depend on the transverse momentum imbalance momentum $\vec \Delta$. The structure  and relative weight of different contributions to the nuclei $_2GPD$ was 
studied in detail in Ref.~\cite{BSW}, where it was shown that only two contributions survive:  the one that corresponds  to DPS1 mechanism and an other corresponding to DPS2.
\par Since our analysis will especially deal with impact parameter $B$ dependence of the cross section,  we find natural to rewrite 
Eq.~(\ref{Eq:DPS_momentum_space}) in coordinate space, introducing the double distributions $D_{p,A}$ which are the Fourier
conjugated of $G_{p,A}$ with respect to $\vec \Delta$. 
In such a representation these distributions admit a probabilistic interpretation and represent the number density of parton pairs with longitudinal fractional momenta $x_1,x_2$, at a relative 
transverse distance ${\vec{b}_\perp}$,  the latter  being the Fourier conjugated to $\vec \Delta$.
\par In the impulse approximation for the nuclei, neglecting possible corrections to factorisation due to the shadowing for large nuclei, and taking into account that $R_A \gg R_p$  for heavy nuclei, we can rewrite the cross section as~\cite{BSW,Strikman:2001gz}
\begin{multline}
\frac{d \sigma^{CD}_{DPS}}{d\Omega_1 d\Omega_2} = \frac{m}{2} 
\sum_{i,j,k,l} 
\sum_{N=p,n} \int d \vec{b}_\perp
\int d^2 B \, 
D^{ij}_{p}(x_1,x_2;\vec{b}_\perp) 
D^{kl}_{N}(x_3,x_4;\vec{b}_\perp)  T_N(B) 
\frac{d \hat \sigma_{ik}^C}{d\Omega_C} \frac{d \hat \sigma_{jl}^D}{d\Omega_D}\,,\\
+ \frac{m}{2} \sum_{i,j,k,l} \sum_{N_3, N_4=p,n}
\int d \vec{b}_\perp
D^{ij}_{p}(x_1,x_2;\vec{b}_\perp) 
\int d^2 B \,
f^{k}_{N_3}(x_3)
f^{l}_{N_4}(x_4)  T_{N_3}(B) T_{N_4}(B) 
\frac{d \hat \sigma_{ik}^C}{d\Omega_C} \frac{d \hat \sigma_{jl}^D}{d\Omega_D}\,.
\label{uno}
\end{multline}
Here $m=1$
if $C$ and $D$ are identical final states and $m=2$ otherwise,
$i,j,k,l = \{q, \bar q, g \}$ are the parton species contributing to the final states $C(D)$.
In Eq. (\ref{uno}) and in the following,
$d \hat{\sigma}$ indicates the partonic cross section for producing the final state $C(D)$, differential in the relevant set of variables, $\Omega_C$ and $\Omega_D$, respectively. 
The functions $f^i$  appearing in  Eq.~(\ref{uno}) 
are single parton densities and the subscript $N$ indicates nuclear  parton distributions.  The double parton diistribution $D_N$ is the double GPD for the nucleon  bound in the nuclei,
once again calculated in the mean field approximation.

\par Partonic cross sections and parton densities do additionally depend on factorization and renormalization scales whose values are set to appropriate combination of the large scales occuring in final state $C$ and $D$. 
\par The nuclear thickness function $T_{p,n}(B)$, mentioned in the Introduction and appearing in Eq.~(\ref{uno}), is obtained integrating the proton and neutron densities $\rho_0^{(p,n)}$ in the nucleus over the longitudinal component $z$
\begin{equation}
\label{thickness}
T_{p,n}(B)=\int dz \rho^{(p,n)}(B,z)\,,
\end{equation}
where we have defined $r$, the distance of a given nucleon from nucleus center, in terms of the impact parameter $B$ between the colliding proton and nucleus, $r=\sqrt{B^2+z^2}$.
Following Ref.~\cite{Alvioli:2018jls}, for the ${}^{208} P_b$ nucleus, the density of proton and neutron is described by a Wood-Saxon distribution
\begin{equation}
\label{Wood-Saxon}
\rho^{(p,n)}(r)=\frac{\rho_0^{(p,n)}}{1+e^{(r-R_0^{(p,n)})/a_{(p,n)}}}\,.
\end{equation}
For the neutron density we use $R_0^n=6.7$ fm and $a_n=0.55$ fm \cite{Tarbert:2013jze}. For the proton density we use $R_0^p=6.68$ fm and $a_p=0.447$ fm \cite{Warda:2010qa}. The $\rho_0^{(p,n)}$ parameters are fixed by requiring that the proton and neutron density, integrated over all distance $r$, are normalized  to the number of the protons and neutrons in the lead nucleus, respectively.
\par As already anticipated, the 
DPS1 contribution, the first term in Eq.~(\ref{uno}), stands for the contribution already at work in $pp$ collisions. 
It depends linearly on the nuclear thickness function $T$ and therefore scales as the number of nucleon in the nucleus, $A$. 
\par The second term, the DPS2 contribution, contains in principle 
two-body nuclear distributions. 
We work here in the impulse approximation, neglecting short range correlations  in the nuclei since their contribution may change the results by several percent only~\cite{Alvioli:2019kcy}.
The latter term is therefore proportional to the product of one-body nucleonic densities in the nucleus, \textsl{i.e.} it depends quadratically on $T$ and parametrically scales as $A^{4/3}$.
\par As we already stated above we shall work here for simplicity in the mean field approximation for the nucleon. In such approximation double GPD has a  factorized form :
\begin{equation}
\label{fact}
D^{ij}_p(x_1,x_2,\mu_A,\mu_B,\vec{b}_\perp) \simeq 
f^{i}_p (x_{1},\mu_{A})
f^{j}_p (x_{2},\mu_{B})\,
\mathcal{T}(\vec{b}_\perp)~,
\end{equation}
where the function $\mathcal{T}(\vec{b}_\perp)$ 
describes the probability to find two partons 
at a relative transverse distance $\vec b_\perp$ in the nucleon and  is normalized to unity. In such a simple approximation, this function does not depend on parton flavour and fractional momenta. Then one may define the so-called effective cross section as
\begin{equation}
\sigma_{eff}^{-1} =  \int d \vec{b}_\perp [\mathcal{T}(\vec{b}_\perp)]^2~,
\label{bprofile}
\end{equation}
which controls the double parton interaction rate.
Under all these approximations the DPS cross section in $pA$ collision can be rewritten as
\begin{multline}
\frac{d \sigma^{CD}_{DPS}}{d\Omega_1 d\Omega_2} = \frac{m}{2} 
\sum_{i,j,k,l} 
\sum_{N=p,n} \sigma_{eff}^{-1}
 f^{i}_{p}(x_1) f^{j}_{p}(x_2)
f^{k}_{N}(x_3) f^{l}_{N}(x_4) 
\frac{d \hat \sigma_{ik}^C}{d\Omega_C} \frac{d \hat \sigma_{jl}^D}{d\Omega_D} \int d^2 B \, T_N(B)\,,\\
+ \frac{m}{2} \sum_{i,j,k,l} \sum_{N_3, N_4=p,n}
f^{i}_{p}(x_1) f^{j}_{p}(x_2)
f^{k}_{N_3}(x_3) f^{l}_{N_4}(x_4) 
\frac{d \hat \sigma_{ik}^C}{d\Omega_C} \frac{d \hat \sigma_{jl}^D}{d\Omega_D}\,
\int d^2 B \, T_{N_3}(B) T_{N_4}(B)\,.
\label{due}
\end{multline}
We find important to remark the key observation that leads to the  second term of Eq.~(\ref{uno}): namely that the $b$ and $B$ integrals practically decouple since the nuclear density does not vary on subnuclear scale~\cite{Strikman:2001gz,Calucci:2013pza,BSW}.
As a result this term does depend on $_2$GPDs integrated over transverse 
distance $b_\perp$, \textsl{i.e.} at $\vec \Delta =0$, for which we assume again mean field approximation:
\begin{equation}
\int d \vec{b}_\perp
D^{ij}_{p}(x_1,x_2;\vec{b}_\perp) \simeq f^{i}_p (x_{1}) f^{j}_p (x_{2})\,.\label{61}
\end{equation}
After integration over 
$b_\perp$ in Eq.~(\ref{bprofile}),
$\sigma_{eff}$ will be the only non-perturbative parameter 
characterising the DPS1 cross section.
We use in our calculation $\sigma_{eff}$ values extracted from experimental analyses of DPS processes in $pp$ collisions. We neglect corrections due to  longitudinal correlations in the nucleon~\cite{BSW}
and any possible dependence of $\sigma_{eff}$ on energy~\cite{BS}.
For the considered final state a number of experimental analyses have extracted its values for $pp$ collisions at $\sqrt{s}$=7 TeV which are reported in the Tab.~(\ref{sigmaeff}).
In our numerical estimates we use the average of those values, $\bar{\sigma}_{eff}=19$ mb.
\setlength{\extrarowheight}{0.2cm}
\begin{table}[t]
\begin{center}
\begin{tabular}{c|c|c}  \hline  \hline
Ref.  & selection &  $\sigma_{eff} [mb] $\\ \hline
\cite{Aaboud:2016dea}  &
$N_{jets} \geq 4$, $p_T^j \geq 20$ GeV , $|\eta_j| \leq 4.4$ & $14.9^{+1.2}_{-1.0}(stat.)^{+5.1}_{-3.8}(syst.)$ \\
& and at least one having $p_T \geq 42.5$ GeV & \\ \hline
\cite{Chatrchyan:2013qza}  & $N_{jets}=4$: two jets with $p_T \geq 50$ GeV & $19.0^{+4.6}_{-3.0}$ \cite{Gunnellini:2014kwa}\\ 
& two jets with $p_T \geq 20$ GeV,  $|\eta_j| \leq 4.7$ &  \\ \hline
\cite{Khachatryan:2016rjt}  & two light jets and two $b$-jets with $p_T \geq 20$ GeV & $23.3^{+3.3}_{-2.5}$ \cite{Gunnellini:2014kwa} \\
& $|\eta_b| \leq 2.4$, $|\eta_j| \leq 4.7$  & \\ \hline
\end{tabular}
\caption{\textsl{Kinematic selection for the $4j$ and $2b2j$ final states adopted in experimental analyses and the corresponding values of extracted $\sigma_{eff}$.}}
\label{sigmaeff}
\end{center}
\end{table}

\par We close this Section by specifying the kinematics and additional settings with which we evaluate Eq.~(\ref{due}).
We consider proton lead collisions 
at a centre-of-mass energy
$\sqrt{s_{pN}}$ = 8.16 TeV. Due to the different energies of the proton and lead beams ($E_p = 6.5$ TeV and $E_{Pb} = 2.56$ TeV per nucleon), the resulting proton-nucleon centre-of-mass is boosted with respect to the
laboratory frame by $\Delta y = 1/2 \, \ln E_p/E_N$ = 0.465 in the proton direction, assumed to be at positive rapidity. 
Therefore jets rapidities, in this frame, 
are given by $y_{CM}=y_{lab}-\Delta y$.
All calculations are based on proton-nucleon centre-of-mass rapidities. 
\par All the relevant DPS and SPS cross sections contributing to the $4j$ and $2b2j$ final states have been calculated  to leading order with \texttt{ALPGEN}~\cite{Mangano:2002ea}. 
Jet cross sections are obtained by identifying final state partons as jets, as appropriate for a leading order calculations.  
\par We use~\texttt{CTEQ6L1} leading order free proton parton distributions ~\cite{Pumplin:2002vw}.
Nuclear effects on the cross sections are estimated by using ~\texttt{EPS09} nuclear parton distributions~\cite{Eskola:2009uj} in separate simulations. They are found to reduce the dijet cross sections 
less than 1\% for $p_T^j>20$ GeV and are neglected. We also mention that 
dijet cross sections are, to very good accuracy, the same on target protons or neutrons, 
so no isospin corrections is applied.

\section{Results : $4j$}
\label{4j}
In this Section we present results for the inclusive production
of, at least, four light jets. Two leading jets are requested to have 
$p_T^{j_1,j_2}>50$ GeV, the subleading ones $p_T^{j_3,j_4}>20$ GeV 
and $|y_j^{lab}|<4.7$. 
Different cuts on the leading and subleading jets 
are enforced to facilitate the pairing for the DPS selection. Both for the DPS and the SPS mechanisms we require the interparton distance in the $\eta-\phi$ plane
\begin{equation}
\Delta R_{ij} = \sqrt{(\eta_i - \eta_j)^2+(\phi_i-\phi_j)^2}    
\end{equation}
to be $\Delta R_{ij}>0.7$, where 
$i$ and $j$ stands for a generic light jets $(i,j=1 \ldots 4, i \neq j)$. In the DPS cross section we set the symmetry factor $m=2$ when the subleading jets have $20<p_T^{j_3,j_4}<50$ GeV and 
$m=1$ if $p_T^{j_3,j_4}>50$ GeV. 
The factorization and renormalization scales
are fixed to $\mu_F=\mu_R=\sqrt{\sum_{j}^{Njet} p_{T,j}^{2}}$, where $N_{jet}=2$ in DPS and  $N_{jet}=4$ in SPS. All the calculations are 
performed with \texttt{ALPGEN}~\cite{Mangano:2002ea}.
\setlength{\extrarowheight}{0.2cm}
\begin{table}[t]
\begin{center}
\begin{tabular}{cccccccc}  \hline  \hline 
 & \hspace{0.4cm} DPS1 \hspace{0.4cm} & \hspace{0.4cm} DPS2 \hspace{0.4cm}  
 & \hspace{0.4cm} SPS \hspace{0.4cm} & \hspace{0.4cm} Sum \hspace{0.4cm} 
 & \hspace{0.4cm} $\sigma(4j)/\sigma(2j)$ \hspace{0.4cm} &  $f_{DPS1}$  &  \hspace{0.4cm}  $f_{DPS2}$ \\ 
$4j$ &  [$\mu$b] & [$\mu$b] & [$\mu$b] & [$\mu$b] &  & & \\ \hline
$p_T^{j_3,j_4}>20$ GeV &  26.0 &  72.2 & 170.9 & 269.2  &   0.15  &  0.13 &   \hspace{0.4cm}   0.27  \\
$p_T^{j_3,j_4}>25$ GeV &  10.8 &  30.2 & 92.9 & 133.9  &   0.07  &  0.10 &   \hspace{0.4cm}   0.22  \\   
$p_T^{j_3,j_4}>30$ GeV &  5.1 & 14.3 & 51.4 & 70.9  &   0.04  &  0.09 &   \hspace{0.4cm}   0.20   \\
\hline
\end{tabular}
\caption{\textsl{Predictions for
$4j$ DPS and SPS cross sections in $pA$ collisions in fiducial phase space, for different cuts on jets transverse momenta.}}
\label{4j:cs}
\end{center}
\end{table}
\begin{figure}[t]
\includegraphics[scale=0.60]{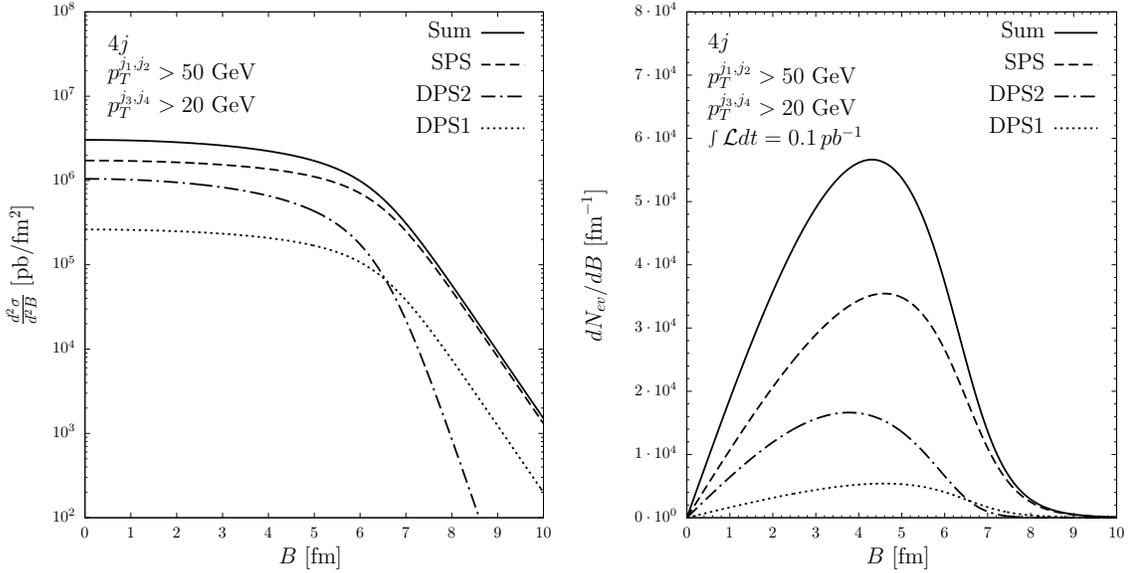}
\caption{\textsl{Differential cross section as a function of $B$ 
for the various contributions to the $4j$ final state (left). 
Expected number of events for the various contributions assuming $\int \mathcal{L} \mbox{dt}= 0.1 \, \mbox{pb}^{-1}$ (right).}}
\label{plot:Nev_4j}
\end{figure}
\par We report in Tab.~\ref{4j:cs} 
the various contributions to the $4j$ fiducial cross section for three different tranverse momentum cuts on the subleading jets. In the last three columns
we report the ratio between the 
4 jets (SPS+DPS) over 2 jets (with $p_T>50$ GeV) cross section, the DPS1 fraction $f_{DPS1}$ calcualted as DPS1 over (DPS1+SPS) cross section ( for easy reference to $pp$ collisions) and the DPS2 fraction $f_{DPS2}$, calculated as  
DPS2 over (DPS1+DPS2+SPS) cross section. 
In general we observe a large contributions from DPS2, which reaches 27\% of 4 jets cross section for $p_T^{j_3,j_4}>20$ GeV.
We present in the left panel of Fig.~(\ref{plot:Nev_4j}) 
the various contributions to the cross sections differential in $B$ 
and the right panel the expected number of events  assuming $\int \mathcal{L} \mbox{dt}= 0.1 \, \mbox{pb}^{-1}$, a value in line with data recorded in 2016 $pA$ runs.
Exploiting the different dependence on $T$ of the various contributions,
we may use the strategy put forward in Ref.~\cite{Alvioli:2019kcy} to separate the DPS2 contribution. For this purpose we evaluate
the number of events  integrating 
Eq.~(\ref{e1}) in the $i$-bin specified by the the bin-edge values $T_i$ and $T_{i+1}$:
\begin{equation}
N_{ev}(T_{i},T_{i+1})=\int
d^2 B \; \frac{d^2 \sigma_{pA}}{d^2 B} \; \Theta\Big(T_A(B)-T_{i}\Big) \; \Theta\Big(T_{i+1}-T_A(B)\Big)
\end{equation}
and then we consider the ratio  $R_{4j}$ 
between the total number (DPS+SPS) of $4jets$ events over those for
dijet production (with $p_T>50$ GeV) as a function of $T_A(B)$:
\begin{equation}
R_{4j}(T_{i},T_{i+1}) = N_{4j}(T_{i},T_{i+1})/ N_{2j}(T_{i},T_{i+1}).
\label{r4j}
\end{equation}
In such a ratio, $N_{2j}$ is linear in $T_A(B)$, as well as the SPS background and the DPS1 mechanisms which both contribute to $N_{4j}$.
In absence of the quadratic DPS2 contribution, 
such a ratio would be a constant. 
Its presence, on the other hand, will induce
a linear increase of the ratio as a function of $T$, 
and the DPS2 magnitudo will determine its slope.
\begin{figure}[t]
\includegraphics[scale=0.60]{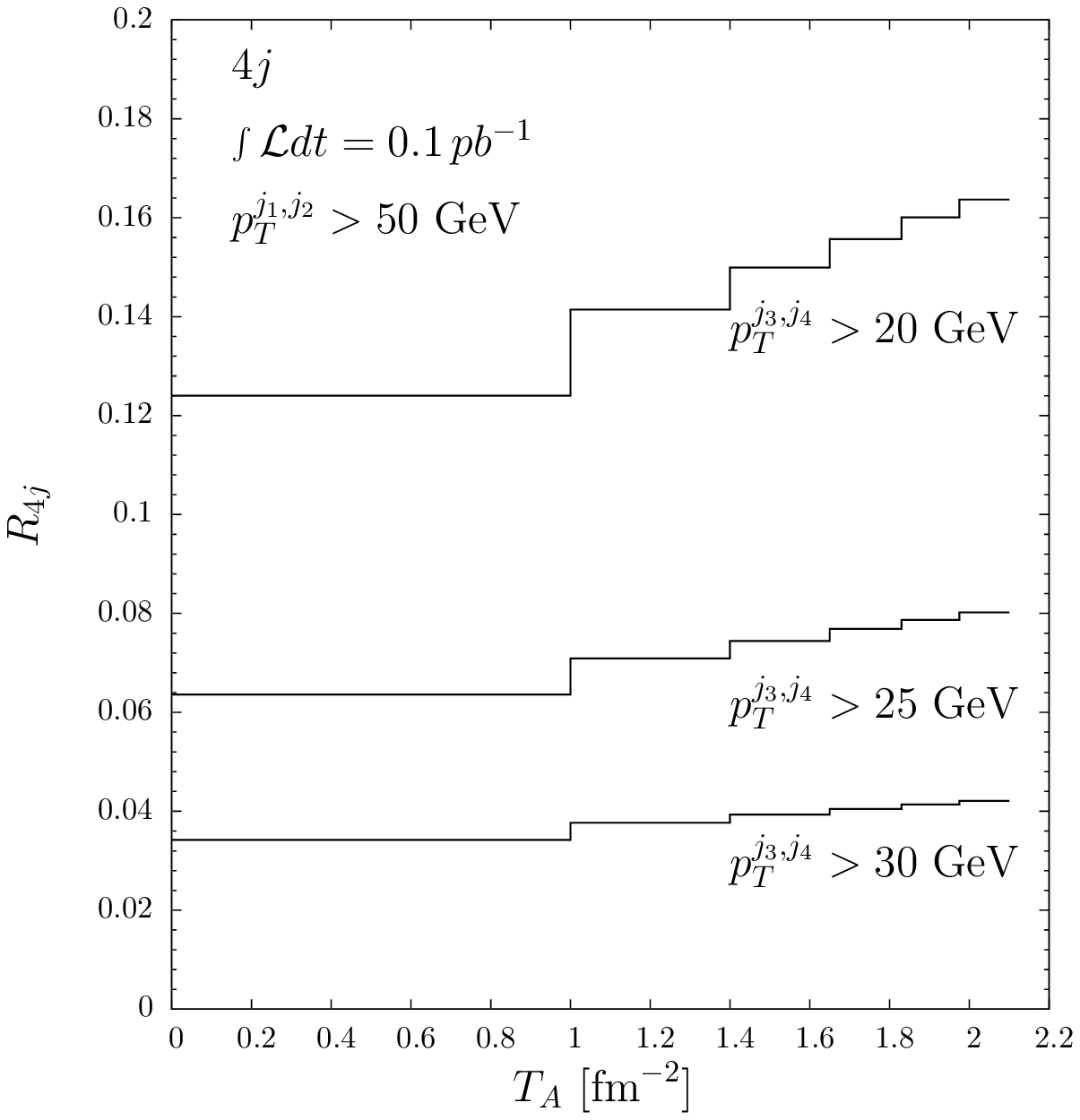}
\includegraphics[scale=0.60]{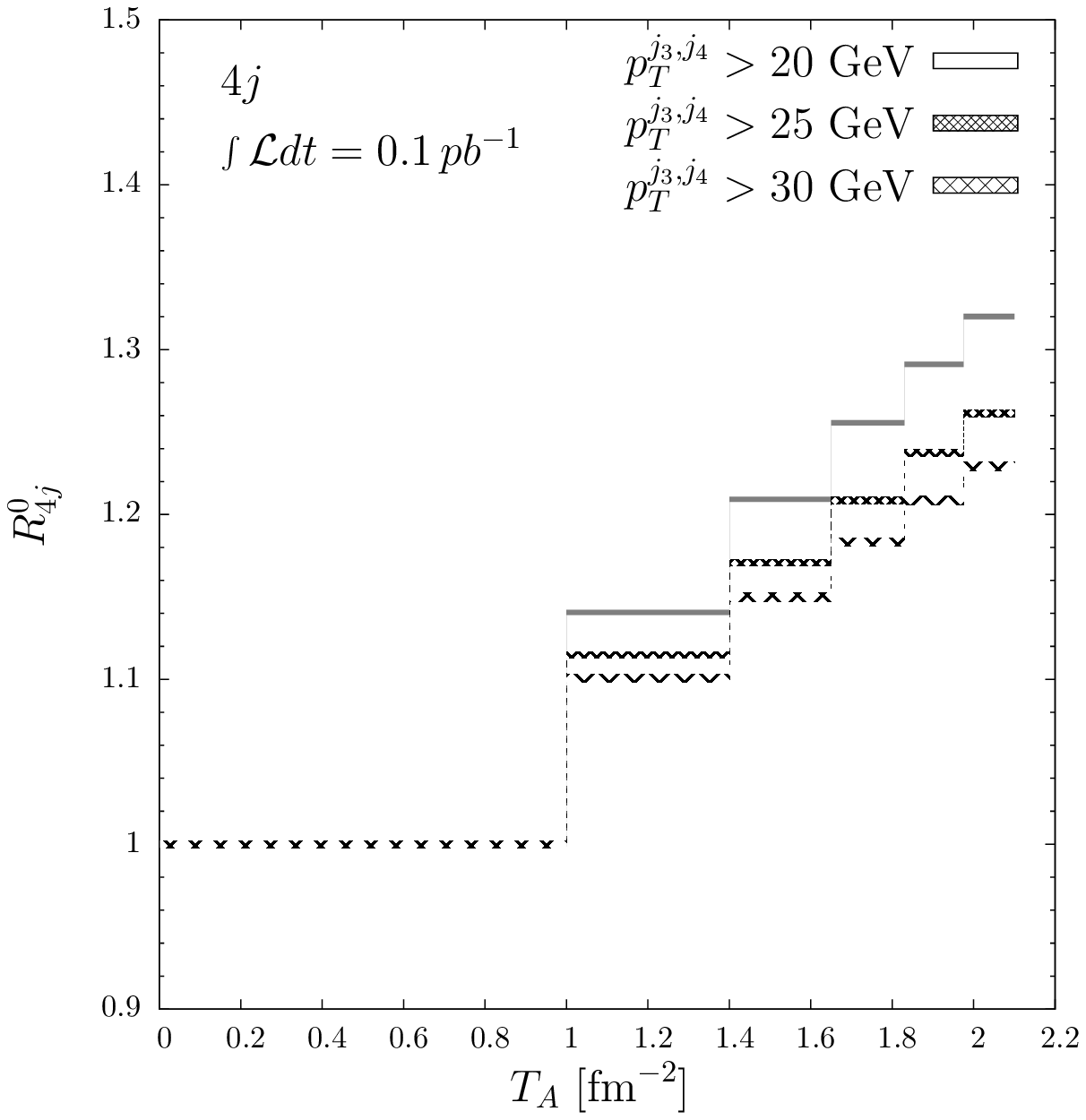}
\caption{\textsl{The ratio in Eq.~(\ref{r4j}) (left) and double ratio in Eq.~(\ref{r4j0})
(right) integrated in bins of $T_A(B)$.  Predictions are shown for three different cuts on jet transverse momenta.}}
\label{plot:4j}
\end{figure}
\par The resulting distribution is presented in the left panel of Fig.~(\ref{plot:4j}) for different values of jet transverse momenta cut off and integrated in bins of $T$, chosen to evenly distribute the number of events. The rise of the slope is related to fast rise of the dijet cross sections entering the DPS2 estimation as the cuts on jet transverse momenta are decreased.
Our calculations were done to the LO (Leading Order) in strong coupling. 
Therefore it is natural to ask for the stability of the ratio in Eq.~(\ref{r4j}).
The role of higher order corrections for the $4j$ final state 
has been investigated in a number of papers and corrections 
has been found to be large~\cite{Badger:2012pf,Bern:2011ep}.
In order to partially overcome this problem, we form double ratio
\begin{equation}
R_{4j}^0(T_{i},T_{i+1}) = \frac{N_{4j} (T_{i},T_{i+1})}{N_{2j}(T_{i},T_{i+1})}
\Bigg( \frac{N_{4j} (T_0,T_1)}{N_{2j}(T_0,T_1)} \Bigg)^{-1}
\label{r4j0}
\end{equation}
\textsl{i.e.} we normalize it to the first bin with 
$T_0=0$ fm$^{-2}$ and $T_1=1$ fm$^{-2}$.
The resulting distribution is presented in the right panel of Fig.~(\ref{plot:4j}). Assuming that statistical errors follow a Poissonian distribution, the associated error is derived from
 the expected number of events. Our results indicate 
that, within these errors estimates, the departure from a constant behaviour can be unambiguously appreciated and the DPS2 contribution disentangled already from data of 2016 $pA$ runs, modulo the experimental issues in studying the most peripheral events.

\section{Results : $2b 2j$}
\label{2b2j}
We consider in this Section a special class of the former 
process in which the second scattering produces a 
$b\bar{b}$ heavy-quark pair. Experimental results 
for this final state are reported in Ref.~\cite{Khachatryan:2016rjt}.
Light and heavy quarks jet are all requested to have $p_T>20$ GeV. 
Additionally light jet are requested to have $|\eta_j^{lab}|<4.7$ and heavy quarks jets $|\eta_b^{lab}|<2.4$. For this final state, the symmetry factor in the DPS cross sections is set to $m=2$. The additional heavy quark tagging facilitate 
the pairing in the DPS selection. Both for DPS and SPS mechanisms 
we set $\Delta R_{ij}>0.7$ where both index runs over light and heavy quarks jets. 
\setlength{\extrarowheight}{0.2cm}
\begin{table}[t]
\begin{center}
\begin{tabular}{cccccccc}  \hline  \hline 
 & \hspace{0.4cm} DPS1 \hspace{0.4cm} & \hspace{0.4cm} DPS2 \hspace{0.4cm}  
 & \hspace{0.4cm} SPS \hspace{0.4cm} & \hspace{0.4cm} Sum \hspace{0.4cm} 
 & \hspace{0.2cm} $\sigma (2b2j)/\sigma (2j)$ \hspace{0.4cm} &  $f_{DPS1}$  &  \hspace{0.4cm}  $f_{DPS2}$ \\  
$2b2j$  &  [$\mu$b] & [$\mu$b] & [$\mu$b] & [$\mu$b] & $\cdot 10^{-4}$ &  &\\ \hline
$p_T^{b,j}>20$ GeV &  2.2 &  6.2 & 13.0 &  21.4  &  3.0 &    0.15    &   \hspace{0.4cm}  0.29 \\
$p_T^{b,j}>25$ GeV &  0.4 &  1.2 & 4.7 & 6.4     &  2.1 &    0.09    &  \hspace{0.4cm}  0.19 \\   
$p_T^{b,j}>30$ GeV &  0.1 &  0.3 & 1.9 & 2.3     &  1.6 &    0.06    &   \hspace{0.4cm}  0.13 \\
\hline
\end{tabular}
\caption{\textsl{Predictions for
$2b2j$ DPS and SPS cross sections in $pA$ collisions in fiducial phase space for different cuts on jets transverse momenta.}}
\label{2b2j:cs}
\end{center}
\end{table}
The factorization and renormalization scales
are fixed to $\mu_F=\mu_R=\sqrt{\sum_{j}^{Njet} m_{T,j}^{2}}$, where $N_{jet}=2$ in DPS and  $N_{jet}=4$ in SPS, being $m_{T,j}=\sqrt{m_{j}^{2}+p_{T,j}^{2}}$, the transverse mass of jet $j$. All the calculations are 
performed with \texttt{ALPGEN}~\cite{Mangano:2002ea}.
\begin{figure}[t]
\includegraphics[scale=0.60]{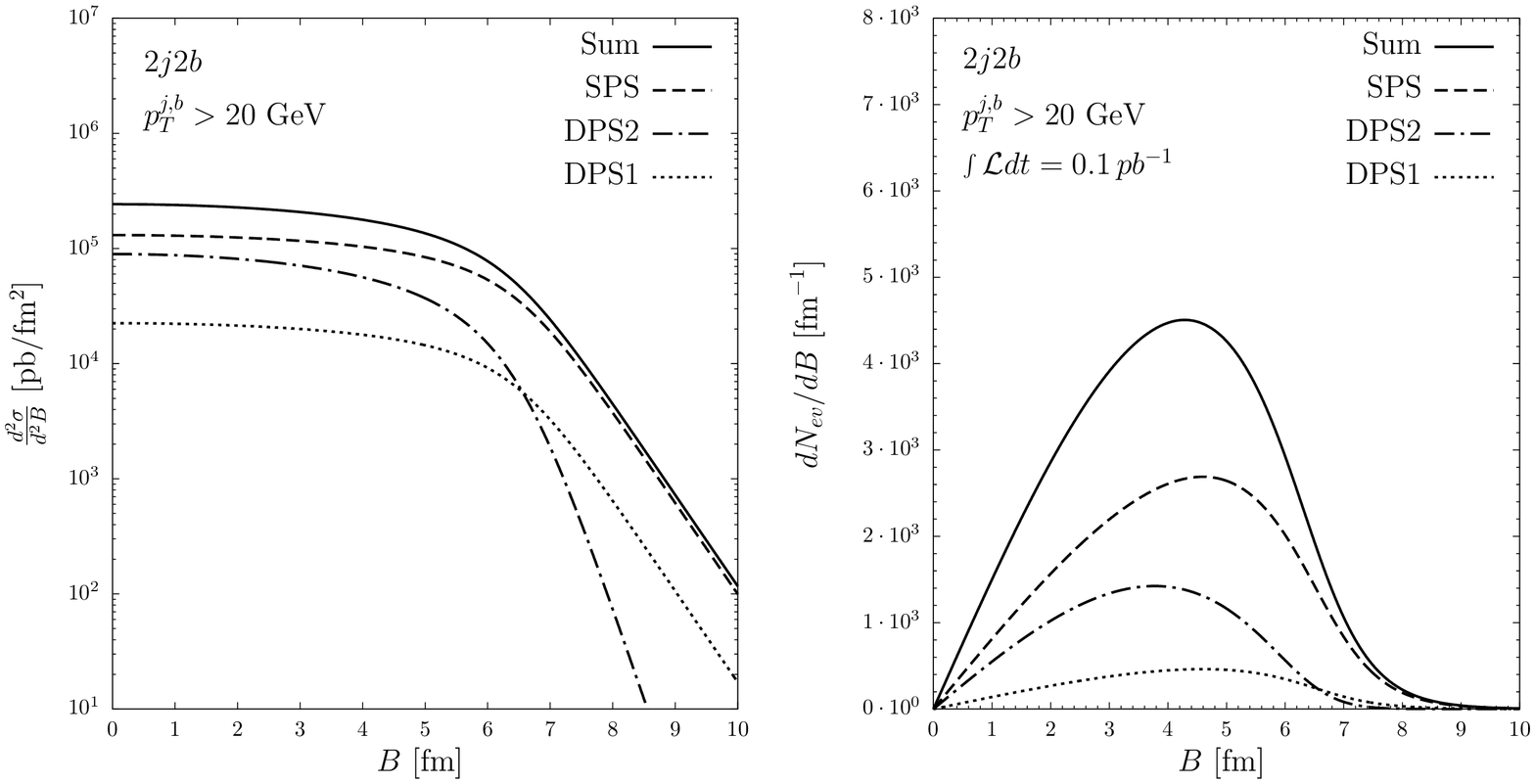}
\caption{\textsl{Differential cross section as a function of $B$ 
for the various contributions to the $2b2j$ final state (left). 
Expected number of events for the various contributions assuming $\int \mathcal{L} \mbox{dt}= 0.1 \, \mbox{pb}^{-1}$ (right).}}
\label{plot:Nev_2b2j}
\end{figure}

\begin{figure}[t]
\includegraphics[scale=0.60]{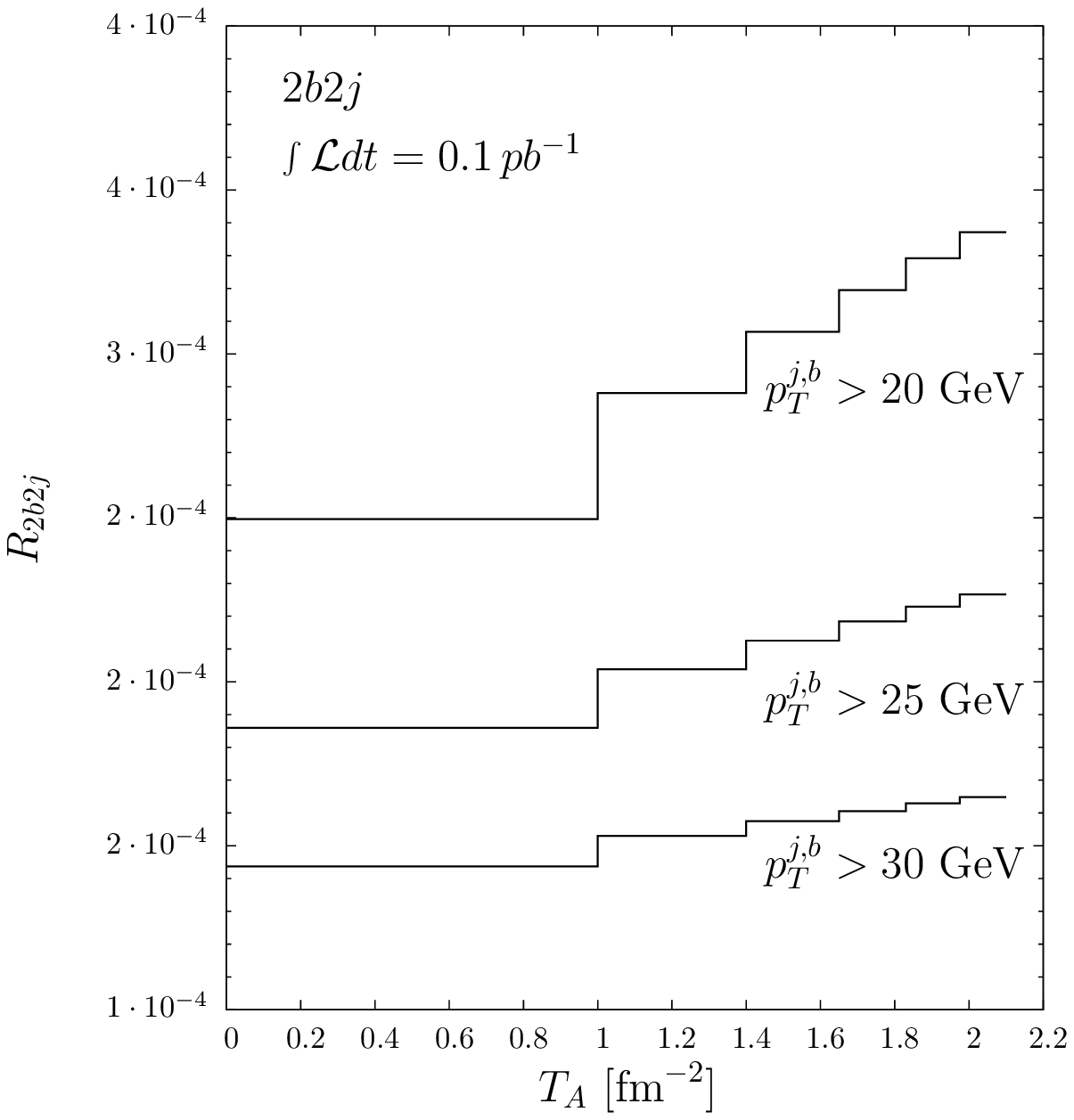}
\includegraphics[scale=0.60]{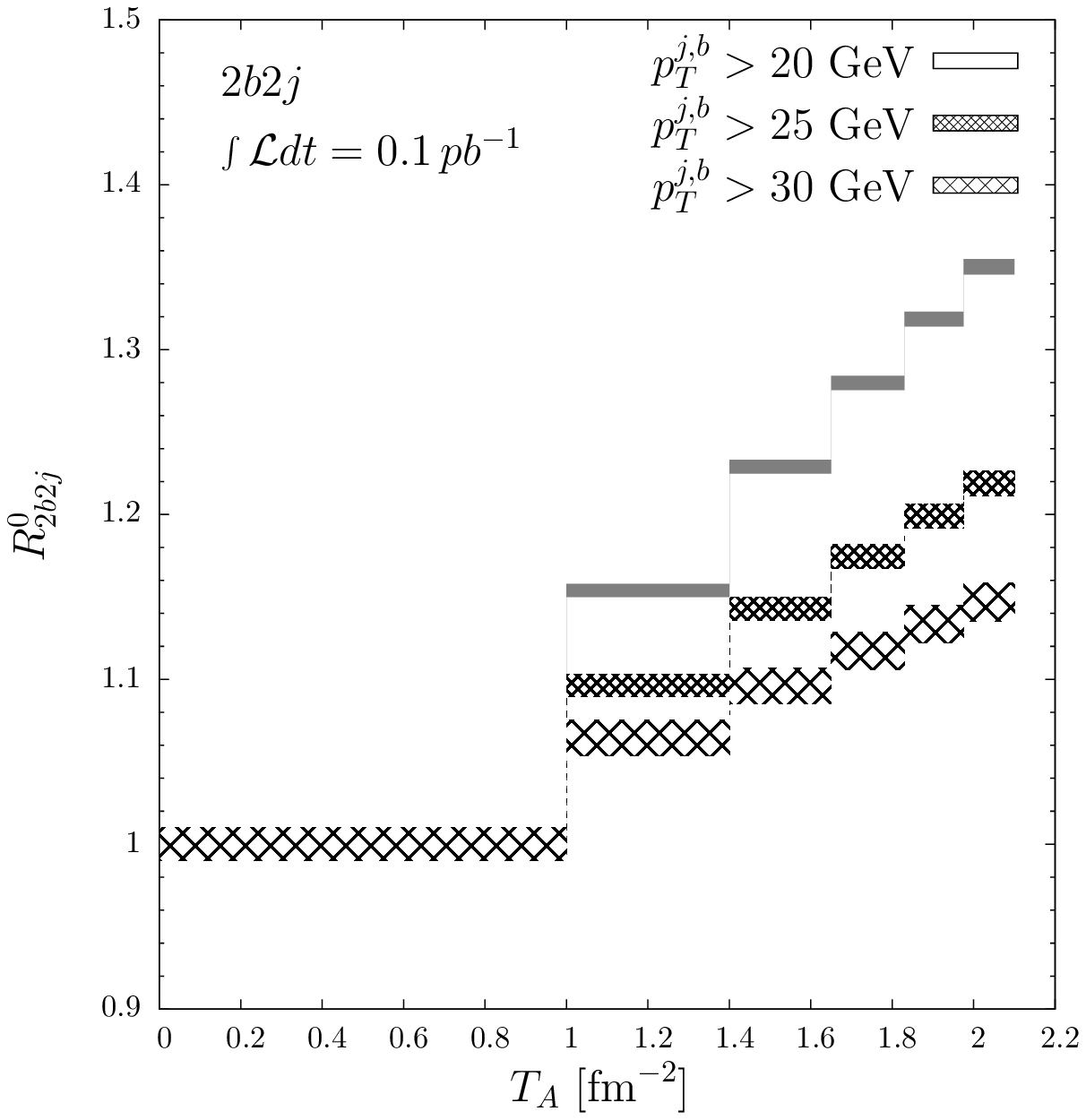}
\caption{\textsl{The ratio in Eq.~(\ref{r2b2j}) (left panel) and double ratio in Eq.~(\ref{doubler:r2b2j}) (right panel) integrated in bins of $T_A(B)$. Predictions are shown for three different cuts on jet transverse momenta.}}
\label{plot:2b2j}
\end{figure}
We report in Tab.~\ref{2b2j:cs} 
the various contributions to the $2b2j$ fiducial cross section for three different transverse momentum cuts on the jets. 
In general we observe a large contributions from DPS2, which reaches 29\% for $p_T>20$ GeV.
We present in the left panel of Fig.~(\ref{plot:Nev_2b2j}) 
the various contributions to the cross sections differential in $B$ 
and the right panel the expected number of events  assuming $\int \mathcal{L} \mbox{dt}= 0.1 \, \mbox{pb}^{-1}$.
As in the previous Section, we consider the ratio  $R_{2b2j}$ 
between the total number of $2b2j$ events (DPS+SPS) over those for
dijet production as a function of $T_A(B)$:
\begin{equation}
R_{2b2j}(T_{i},T_{i+1}) = N_{2b2j} (T_{i},T_{i+1})/ N_{2j}(T_{i},T_{i+1}).
\label{r2b2j}
\end{equation}
\par The resulting distribution is presented in the left panel of Fig.~(\ref{plot:2b2j}) for different values of jet transverse momenta cut off and integrated in bins of $T$. The rise of the slope is related to fast rise of the dijet cross sections entering the DPS2 estimation as the cuts on jet transverse momenta are decreased.
As shown in Tab.(3) of Ref.~\cite{Khachatryan:2016rjt}, the comparison of various 
theoretical predictions with $2b2j$ data reveal 
substantial agreement with NLO predictions but LO prediction suffers from large higher order corrections. Such a results are confirmed also by \texttt{ALPGEN} prediction which returns a cross section 0.6 times smaller than data~\cite{Khachatryan:2016rjt}. In order to partly  mitigate these effects, we form the double ratio 
\begin{equation}
R_{2b2j}^0 (T_{i},T_{i+1}) = \frac{N_{2b2j} (T_{i},T_{i+1})}{N_{2j} (T_{i},T_{i+1})}
\Bigg( \frac{N_{2b2j} (T_0,T_1)}{N_{2j}(T_0,T_1)} \Bigg)^{-1}
\label{doubler:r2b2j}
\end{equation}
\textsl{i.e.} we normalize it to the first bin ($0<T<1$).
The resulting distribution is presented in the right panel of Fig.~(\ref{plot:2b2j}). The associated error is calculated from 
the expected number of events, assuming a Poissonian  distribution 
for statistical errors. 
\par Our results indicate 
that, although  with lesser significance with respect to the four-jet case, the departure from a constant behaviour can be unambiguously observed also in this final state. As already observed in the 
$4j$ case, lowering the cut on the jet transverse momenta increases the sensitivity to a non constant behaviour of $R_{2b2j}^0$.  

\section{Conclusions}
\label{conclusive}
\indent In this paper we have calculated 
DPS cross sections for  double dijet final states produced in $pA$ collisions at the LHC, as well as the corresponding SPS backgrounds. 
Relying on the experimental capabilities to 
correlate centrality with impact parameter $B$ of 
the proton-nucleus collision, 
we have presented a strategy to extract the so-called DPS2 contributions, pertinent to $pA$ collisions.
With this respect 
the $4j$ and $2b2j$ final states has large enough cross sections to  allow the use  of the method \cite{Alvioli:2019kcy} 
to disentangle  Leading Twist + DPS1 contributions from the DPS2 contribution, which is the main interest of this paper, already with data accumulated in 2016 $pA$ run.

\begin{acknowledgments}
\noindent
The authors would like to thank  M. Strikman for reading the manuscript and for many useful discussions.  We also thank A. Milov 
for many useful comments.
The work was supported by Israel Science  Foundation  under  the  grant  2025311.
The diagram in this paper has been drawn with Jaxodraw package version 2.0 \cite{Binosi:2008ig}.
\end{acknowledgments}

  \end{document}